\documentclass[aps,pre,twocolumn,groupedaddress,showkeys]{revtex4}

\usepackage{url}
\usepackage{amssymb}
\usepackage{amsmath}
\usepackage{color}
\usepackage{graphicx}
\usepackage{verbatim}
\usepackage[            
pdftex,                 
pdfborder= 0 0 0,
a4paper,                
colorlinks=true,        
citecolor=green,        
linkcolor=cyan,         
menucolor=blue,         
urlcolor=magenta,       
bookmarksopen=true
]{hyperref}

\begin{document}

\title{AGENT-BASED SOCIAL PSYCHOLOGY: FROM NEUROCOGNITIVE PROCESSES TO SOCIAL DATA\footnote{Preprint of an article submitted for consideration in ACS 2011 $\copyright$World Scientific Publishing Company \url{http://www.worldscinet.com/acs/} }}

\author{Nestor Caticha}
\affiliation{Dep. de Fisica Geral \\
              Instituto de F{\'\i}sica, Universidade de S\~ao Paulo \\
	      CP 66318, 05315-970, S\~ao Paulo - SP, Brazil  \\
nestor@if.usp.br}

\author{Renato Vicente}
\affiliation{Dept. of Applied Mathematics \\
	      Instituto de Matem{\'a}tica e Estat{\'\i}stica, Universidade de S\~ao Paulo \\
	      05508-090, S\~ao Paulo - SP, Brazil \\
rvicente@ime.usp.br}

\begin{abstract}
Moral Foundation Theory  states that groups 
of different observers may rely on  partially dissimilar sets of moral foundations, 
thereby reaching different moral 
valuations. The use of  functional imaging 
techniques  has revealed  a spectrum of cognitive styles with respect to the differential handling of
 novel or corroborating information that is correlated to political affiliation.
 Here we characterize the collective behavior of an agent-based model 
whose inter individual interactions due to information exchange in the form of opinions
are in qualitative agreement with experimental neuroscience data. 
The main conclusion derived connects the existence of diversity in the
 cognitive strategies and  statistics of the sets of moral foundations and suggests
 that this connection arises from interactions between agents. Thus a simple interacting 
agent model, 
whose interactions are in accord with empirical data on conformity and  learning processes, presents
 statistical signatures consistent with moral judgment patterns 
 of conservatives and liberals as obtained by survey studies of social psychology. 
\end{abstract}


\maketitle

\section{Introduction}

The proponents of Moral Foundation Theory (MFT) \cite{Haidt1}  have identified at least five moral foundations or 
dimensions that, potentially, are universally present in humans. These dimensions are manifested
 in different manners, not only across time and cultures, but also within a society.
 Individuals with different attributions of the relative importance of the dimensions
 will be led to fundamental misunderstanding of moral motivations of each other. Within
 the MFT, extensive empirical support \cite{Graham,Haidt2,Haidt3} has been gathered for the fact that the use of 
different subsets of moral foundations by groups is significantly correlated with a 
scale characterizing the group along the political spectrum.  The subsets are
 such that liberals tend to rely more strongly on aspects relating to (a) harm/care 
and (b) fairness/reciprocity. Conservatives rely on these aspects but not as much. 
In addition they also regard as important (c) in-group loyalty, (d) authority/respect
 and (e) purity/sanctity, to a larger extent than liberals \cite{Haidt1}.

At the scale of individuals, empirical evidence supports that: 1. moral opinions are to a 
large extent emitted automatically, that is, people, as a first approximation, 
are intuitionists \cite{Greene2,Greene3,Greene,Haidt2}; 
2. there is a psychological cost of dissent, with humans trying to 
attain social conformity modulated by peer pressure \cite{Asch,Eisenberger,Sherif,Somerville}; 
3. conformity is learned from interactions within a social network \cite{Klucharev}; 
 4. individual cognitive strategies may differ 
with respect to the relative sensitivity to learning from novel information as compared 
to reinforcing habitual responses \cite{Amodio}.

Despite the growing body of experimental evidence accumulated over the last decade,   explicit 
connections between this new empirical evidence on individual behavior and social phenomena (or between micromotives and macrobehavior 
\cite{Schelling}) still are relatively unexplored. Since the work on sociophysics by Galam \cite{Galam82,Galam86},  
the statistical mechanics community has already  addressed this aggregation  problem in social systems 
\cite{Castellano,Durlauf,Galam,Galam91,Roehner,Weidlich}\footnote{That methods from physics can be used in sociology is not
 particularly surprising, as these methods were actually brought to physics in the 19th century by James Clerk Maxwell when he was inspired 
by the historian Henry  Buckle's account of Adolphe Quetelet's statistical approach \cite{Quetelet} to social science (\cite{Maxwell} p. 438).}. 
However, this research has mainly focused on the study of simplified scenarios based on common sense suppositions that result in models that are interesting {\it per se}.
 Some recent works have exemplified a different direction by postulating  reasonable  inter-individual interactions and trying to predict \cite{GalamQQ} or explain
 \cite{Bouchaud}  empirically observed aggregate behavior. We believe that the program of  building models explicitly based on the empirical evidence  that is now 
available \cite{Fowler} is worth pursuing. 

We do not believe that any stylized model can provide an all encompassing and precisely quantitative description of  human nature.
Our general goals in this work are instead much more modest and can be stated as follows. Our first goal is to  provide a  mathematical model that is capable of connecting in the same  
framework empirical evidence from processes at different scales. We also would like to have a framework that is capable of instigating the formulation of new theoretical and experimental questions. In 
particular, we propose a model with agents consistent with empirical evidence and study its aggregate behavior by employing the approach and numerical techniques of statistical mechanics.
 We then show that this aggregate behavior predicts  that a  well defined feature is expected to be observed in the data we are considering. We then verify the consistency of our predictions and propose 
a new interpretation to empirical evidence within this framework that can be qualitatively tested against new data sets in the future. We
 insist that such a general model is only expected to provide qualitative predictions of limited scope and emphasize our belief that a 
good model should make testable predictions but should not explain too much.

In the following sections we give details on the empirical evidence we consider relevant, introduce our modeling approach, 
present and discuss the results obtained. A brief summary of methods employed is provided as an appendix.

\section{Empirical evidence for the model}
In building our model we have tried to incorporate in a stylized manner  empirical evidence. 
In this section we describe what we believe to be 
essential empirical observations  and which model structures they suggest.

\subsection{Moral theories and the automaticity of moral judgments}

Philosophers have struggled with the problem of conceptualizing morality since antiquity.  Although philosophical theories tend to be of normative 
character they can be regarded as a starting point in considering a possible scientific approach for morality as a social phenomenon. 
Three theories are of particular interest to our discussion: virtue theory, deontology  and utilitarianism \cite{Broad,Joseph}.

 According to \cite{Casebeer,Joseph},  utilitarianism  proposes that moral judgments should be based on the consequences resulting from them.
Only actions that maximize social happiness and minimize pain should be taken. In the deontological view only actions that could be universally 
adopted without violating anyone's rights should be pursued. Virtue theory takes into account the intrinsic limitations of human nature and
 states that morality is concerned with maximizing virtues and minimizing vices. Each view of morality presupposes cognitive loads that can 
be experimentally verified. While utilitarianism and deontology concentrate moral decisions on higher cognitive functions in the prefrontal 
and sensorial brain regions, virtue theory proposes the coordinated functioning of these areas with others associated to the processing of 
emotions.  Data gathered in the last decade favors a combination of the three views with  preponderance of a mode  that is closer to virtue 
theory \cite{Greene,Damasio}. Actually ample research points in the direction that moral opinions are mostly formed with very 
few recourse to utilitarian (or consequentialist) reasoning. 

Evidence supports that moral violations elicit strong negative responses that activate the socio-emotional structures 
of the brain (e.g. medial prefrontal cortex and posterior cingulate cortex)\footnote{For a  discussion of the emotional components
of moral intuitions and their neural substrates see \cite{Moll} and 
\cite{Woodward}, see \cite{Greene,Greene2} for fMRI evidence and  \cite{Haidt2} for psychological tests.}. But emotions are not the only component 
behind every moral judgment  as experimental observation  also suggests 
 that these automatic negative responses can be overcome by a more utilitarian mode by recruiting cognitive areas in the prefrontal cortex, in particular, 
when difficult personal dilemmas with important social consequences are involved \cite{Greene3,Damasio,Pizarro}.

In this work we simplify by only considering the moral grading of  statements 
and not the comparison and subsequent choice between different
possibilities and their consequences. Therefore we keep from entering a discussion 
between deontological or consequentialist \cite{Broad} moral theories which might
guide the modeling of decisions and  choices associated to moral dilemmas. We start by supposing as a first approximation that 
socio-emotional intuitions predominate and  that moral grading is in fact automatic.  

\subsection{Moral foundations}
Human culture and values are markedly diverse. Nevertheless, this diversity seems to emerge from innate universals. Modern research in cultural anthropology 
\cite{Shweder}, primatology \cite{DeWaal} and evolutionary psychology \cite{Haidt1,Katz} suggests that morality may be parsed into a small number 
of basic intuitions. Haidt and collaborators \cite{Graham,Haidt1,Joseph,Haidt3} reviewed the literature to identify five 
candidates to innate moral intuitions (or foundations) associated to:  care, fairness (classified as individualizing foundations), loyalty, authority and purity (binding foundations).  This 
set of innate moral foundations could have coevolved with culture due to adaptive challenges
 primate populations have been subjected to in their evolutionary history \cite{Boyd,Haidt1}. This innateness, however, does not imply  moral judgments
 that are rigid or genetically determined. What is considered to be a virtue or a vice in a given society at a given time depends instead on learning and imitation 
in a social environment. This plasticity from an initial draft is the key to understand how diversity can be universality-bound 
\cite{Graham,Marcus}. In our model we introduce moral foundations as dimensions in an abstract moral state space (for a similar suggestion see \cite{Churchland}). 
Five dimensional moral vectors live in this space  and are animated by an adaptation dynamics elicited by social interactions.

\subsection{Reinforcement Learning}

Extensive literature (see \cite{Holroyd} and references therein) suggests the existence of a generic machinery
for error and conflict processing in humans. This adaptive circuit implements a full-fledged reinforcement learning system. 
In this system the basal ganglia processes error information provided by  the spinal cord, the sensorial cortex and by 
areas that were traditionally labeled as the limbic system. This error measure is then converted into a dopamine signal that is
 used to correct responses with the mediation of a strategically connected region known as the Anterior Cingulate Cortex (ACC).

In event related potential (ERP) experiments the processing of social exclusion feelings and social normative conflicts  has been associated to
 error related negativity (ERN) signals with  source located in the ACC. This localization has been  further confirmed by fMRI \cite{Holroyd}. 
Initial activation of the ACC due to conflict has also been associated
 to subsequent alignment to opinions perceived as preponderant in the social group \cite{Klucharev}. These new data corroborate classical behavioral experiments
  on the psychology of conformity  conducted by Sherif \cite{Sherif} and Asch \cite{Asch}.  Such findings suggest a central role to reinforcement learning in the dynamics 
of conformity to social norms. The observation of  amigdala  activation during social conflict \cite{Berns} together with the known association of the ACC activation 
when physical pain is involved  \cite{Eisenberger,Somerville} additionally suggest  that disagreement elicits a psychological cost in humans.

Both the  automaticity of moral judgments and  the psychological cost of disagreement can be represented by using  a reinforcement learning model that is  well established  
in computer science  \cite{Sutton} and statistical mechanics \cite{Engel01}. Within this model a moral judgment is regarded as a classification task. Each agent 
has an internal moral state ${\mathbf{J}}_i$.  At each time step an agent is chosen and its internal state is updated 
to minimize the psychological cost. If there is no noise in the communication, this minimization follows a gradient descent dynamics:
\begin{eqnarray}
\tilde{\mathbf{J}}_i(t+1)&=&\mathbf{J}_i(t)-
\epsilon \nabla_{\mathbf{J}_i(t)}{\cal H},\nonumber\\
\mathbf{J}_i(t+1)&=&\frac{\tilde{\mathbf{J}}_i(t+1)}
{|\tilde{\mathbf{J}}_i(t+1)|}\label{noiseless}
\end{eqnarray}
where $\epsilon$ defines the time scale and ${\cal H}$ represents the social cost, namely, the sum of psychological costs incurred by an agent in a given social network.

\subsection {Cognitive styles}

A recent experiment \cite{Amodio} has shown evidence that
there is a correlation between being a  liberal or conservative
with respect to social issues and the way novel or 
corroborating information is used. The experimental setup consists of the 
measurement of ERPs and concomitant fMRI while participants are exposed to a Go/No-Go 
task. The subjects first habituate to a frequent ``Go'' stimulus. In some rare occasions 
a ``No-Go'' stimulus appears and  a related ERN signal is registered. As expected, a localization
algorithm and simultaneous fMRI  identify the ACC as a source for the conflict signal. Before the 
experimental section, the participants are asked to rate their political orientation from $-5$ for 
``very liberals'' to $+5$ for ``very conservatives''.  A negative  correlation is then found 
between political affiliation and the amplitude of the ERN signal. Liberals exhibit more intense conflict related
activation  of the ACC as compared to habituated response.     

Two other recent studies provide further evidence by associating political behavior and genetic differences affecting dopamine receptors 
known as DRD2 and DRD4 \cite{Fowler1,Fowler2}. Dopamine is a neurotransmitter directly related, among 
several other things, to predictive reward systems that modulate reinforcement learning mechanisms \cite{Schultz}.

We regard these experiments as suggesting that self-declared liberals are more at ease with novel 
information and rely less on corroborating information while self-declared conservatives prefer corroboration 
and are less at ease with novelty.  We use the term corroboration  to signify ``confirming and in accordance to previous opinion''.
This new empirical work concurs with ample literature in  social sciences which have discussed for decades the relation between
cognitive styles and political orientation \cite{Adorno}. Whether this cognitive diversity is due to genetic or cultural   
conditions is beyond our present scope.  

We try to capture some aspect of the information conveyed by this class of empirical results by recurring to models of statistical learning 
\cite{Engel01}. We propose that there are different learning styles according to the balance between novelty seeking and corroboration. 
For that we introduce a  parameter $0\le\delta\le1$ that specifies the amplitude ratio between learning corroborating and learning conflicting
 information and define a learning algorithm interpolating between pure corroboration  learning when $\delta=1$ and pure novelty 
seeking  learning when $\delta=0$. In one social interaction we would then have:
\begin{eqnarray}
\tilde{\mathbf{J}}_i(t+1)&=&\mathbf{J}_i(t)+\epsilon  F(h_i,h_j) \mathbf{x} 
\\
\mathbf{J}_i(t+1)&=&\frac{\tilde{\mathbf{J}}_i(t+1)}
{\mid\tilde{\mathbf{J}}_i(t+1)\mid}, \nonumber
\end{eqnarray}
where the function
\begin{equation}
 F(h_i,h_j) = \left\{ \begin{array}{ll}
                       \delta h_j & \mbox{if $h_ih_j>0$} \\
			  h_j   & \mbox{otherwise}
                      \end{array} \right., 
\label{modulation}
\end{equation}
modulates learning of agent $i$  by comparing its classification  $h_i=\mathbf{J}_i\cdot\mathbf{x}$ of an input $\mathbf{x}$  
with that issued by a social neighbor $h_j$. The response of agent $i$ is then corrected with direction given by the issue  and
sign and amplitude dictated by the information provided by agent $j$.

\subsection {Social influence}

Classical experimental setups by Sherif \cite{Sherif} and Asch \cite{Asch}  demonstrated that 
groups influence individual  beliefs and decisions. In Sherif's experiment subjects are placed in 
a dark room  and asked to judge the displacement of a spot of light without knowing that it is actually stationary. The 
task is repeated a number of rounds with participants either alone or in groups. When in group, individual estimates converge to a group specific norm.  

In Asch's experiment a participant is placed in a room with a group of other people that 
are, without her knowledge, confederates of the study. She is then presented with two cards. One 
with a standard vertical line  and the other with one vertical line the same length of the standard and 
two with different lengths. The participant is then asked to identify which of the lines in the second 
card is  most similar to the standard line in the first card, but this is done after 
all confederates unanimously make the wrong choice. A strong conformist trend is observed, 
with decreased accuracy of individual judgment.   

These experiments are generically consistent with the modern picture of conflict mediated reinforcement 
learning \cite{Klucharev}. However, two aspects of these setups require a closer examination: in the first  
setup participants are anonymous and information ambiguous and in the second information is 
comparatively objective and subjects largely uncategorized.

Modern social psychology defines three types of social influence \cite{Abrams}: informational, normative 
and referent informational. Informational influence is the predominant mechanism for social influence 
in the absence of objective evidence and when no group  identification is present as in Sherif's 
experiment. The expectation of acceptance or punishment by other members in a group leads to 
normative influence observed in Asch's experiment. When group membership is salient referent informational
influence becomes dominant. In this mode individuals seek to be identified as pertaining to a given 
group.  
   
To investigate referent informational influence in \cite{Abrams} Sherif's and Asch's experiments are repeated with the introduction of salient group 
membership. It is observed that if a participant regards herself as part of a different group,
 the conformity effect is greatly diminished. 

In our modeling effort we then suppose that referent informational influence is preponderant
 and, as a starting point, assume the extreme scenario where only in-group conflict leads to conformity effects.
We therefore start by considering the case in which agents are circumscribed to social neighborhoods with
 homogeneous cognitive styles represented by the corroboration/novelty parameter $\delta$. 

\subsection{Social topology}
Social networks have received a great deal of attention during the last decade \cite{Albert}. 
The Internet and social networking sites like Facebook, MySpace or Orkut now make possible to study 
empirically topological and dynamical properties of social graphs. One of the simplest topological 
properties that can be defined is the distribution of node degrees $P(k)$. A growing number of studies 
seems to indicate that many natural networks are well represented by scale-free graphs with the 
tail of the node distribution given by $P(k)\sim k^{-\gamma}$\cite{Albert}. 

To make an informed modeling choice we have searched the literature for  networks representing
social interactions. We have found some illustrative cases. For instance, the node distribution
 of a network of  phone calls has been found to be scale-free with $\gamma=2.1$ in  \cite{Aiello}.
 The network of sexual contacts has been  identified as being scale-free with $\gamma=3.4$ in 
\cite{Liljeros}. By employing a sampling algorithm a  power law with $\gamma=3.4$ has been 
reported for the Facebook \cite{Gjoka}, but this estimate relies on a range of degrees spanning only one decade.
Other study of entire networks instead supports an exponential node distribution in this case
 \cite{Traud}. A third study \cite{Ahn} with samples from three social network websites reports 
scale-free behavior with $\gamma=2$, $\gamma=3.1$  and $\gamma=3.7$, 
respectively for Cyworld, MySpace and Orkut.

We have considered empirical evidence and have applied the simple 
procedure of preferential attachment described by Barab\'asi and Albert \cite{Albert} to generate 
scale-free social networks with $\gamma=3$. We are aware of the fact that clustering  properties  of networks built in such way will differ 
from those found in real social networks, however, we have been able to verify that the particular results we present here are sensitive to the 
degree distribution and qualitatively robust in relation to other topological properties \cite{ViSuJeCa}.

\section{Agent-based model}

\subsection{Combining empirical ingredients}
We introduce a model for an interacting society where agents represent individuals
 that debate moral issues with their social neighbors. In general terms our modeling approach
continues a now established line of research on opinion dynamics \cite{Castellano,Galam,Vicente}. We however, strongly emphasize that 
information exchanges and processing, even though stylized,    should be explicitly linked to the empirical evidence available. 

We start by supposing that the moral state space has $M_D=5$ dimensions so that moral issues may be parsed into these dimensions. We 
simplify by assuming only unit vectors. It is certainly possible that the same results we have reached could have been obtained by assuming
 less (or more) dimensions, however, it would be incompatible with known empirical data that supports the existence of five moral dimensions \cite{Haidt2,Haidt1,Haidt3}.

We  consider that an agent $i$ attributes a moral content for an  issue $\mu$ that may then be represented by a five component unit vector
 ({\it  issue vector}) \cite{Vicente}  $\mathbf{x}_{i\mu} = \mathbf{x}_\mu + \mathbf{u}_{i\mu}$ with $\mu=1,...,P$ and $i=1,\cdots,N$. Here 
$\mathbf{x}_\mu$ represents the average part of the moral parsing and $\mathbf{u}_{i\mu}$  represents an individual part.

We call the  average (normalized) issue $\mathbf{Z}\propto \sum_{i=1}^N \sum_{\mu=1}^P \mathbf{x}_{i\mu}$ the 
{\it Zeitgeist vector},
 which can be regarded as describing the cultural environment and providing a symmetry breaking direction in the moral state space.  Here 
we are not to be concerned with the origin of the {\it Zeitgeist}\footnote{For an essay on the role of the {\it Zeitgeist} in historical explanation see \cite{Forland}. } vector as it results from evolutionary and historical processes 
taking place in time scales that exceed the scope of our simple model. We further simplify the model by assuming that individual components 
are such that $\sum_{i=1}^N \sum_{\mu=1}^P \mathbf{u}_{i\mu}=0$ and are small enough so that they can be disregarded in a first analysis. 
We are also going to assume that the individual components are not correlated over the social network.

The relevant variables to characterize an agent are  suggested by
moral foundation theory. For each agent and  unavailable to other agents, the internal moral state is encoded 
in another unit vector $\mathbf{J}_i$  ({\it moral vector}), also five dimensional, the magnitude of each 
component representing the weight the $i$-th agent gives to a particular moral foundation. 
Unit vectors are used to avoid introducing  the collateral notion that one agent could be more moral than another. 

The automaticity of moral judgment is then represented as a classification task where in an elementary interaction  an agent $i$ gathers
information on the moral classification of her social neighbor $j$ on a given issue $\mu$. This classification is represented by 
a field given by $h_{j\mu}=\mathbf{J}_{j}\cdot\mathbf{x}_{j\mu}$. Its attributes are its sign and magnitude, 
indicating respectively whether the issue is considered morally acceptable ($h_{i\mu}>0$) or not ($h_{i\mu}<0$) and how strongly the agent holds this 
position ($\lvert{h_{i\mu}}\rvert$).

 At this point we make an additional abstraction leap that leads 
to a still simpler model: we suppose that a debate is a more complex interaction that involves multiple issues and multiple agents 
 and it works effectively as if the participants were estimating the {\it Zeitgeist} vector $\mathbf{Z}$. Therefore, the effective 
interaction we are going to consider corresponds to the exchange of fields $h_{j}=\mathbf{J}_{j}\cdot\mathbf{Z}$ between neighbors.

We only consider here  social influence between similar cognitive styles, namely,
agents have homogeneous cognitive styles and interactions are symmetric. To consider the empirical fact that in-group disagreement 
(or being in the minority) elicits a negative brain response 
  we introduce a measure of the psychological cost of disagreement between socially interacting agents $i$  and $j$.
 It is quantified by $V_\delta(h_{i},h_{j})$, a function of their opinions which depends on a parameter $\delta$ that measures 
the different treatment of corroborating or novel opinions. 

Reinforcement learning  can be recast in its off-line version \cite{Engel01} as the process of 
seeking a minimum in a given cost landscape. Along this line we assume that moral vectors 
 $\mathbf{J}_i$ evolve by decreasing the psychological cost under communication through a noisy channel.
  The social cost $\mathcal{H}$ is defined by summing $V_\delta$ over all pairs $(i,j)$ of interacting agents:
\begin{equation}
 \mathcal{H}(\{\mathbf{J}_i\})=\sum_{(i,j)} V_\delta(h_{i},h_{j}).
 \label{social_cost}
\end{equation}
 It depends on $\{\mathbf{J}_i\}$, the
 configuration of the society, and on the cultural environment, given by the {\it Zeitgeist} vector. 

\begin{figure}[th]
\hspace*{-0.5cm}\includegraphics[width=9cm]{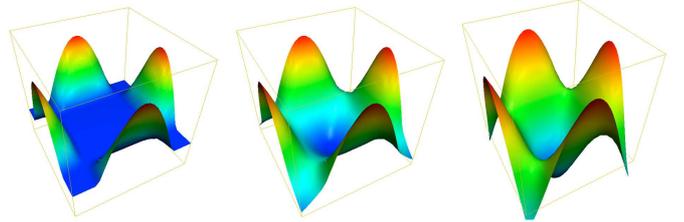}
\vspace*{8pt}
\caption{Psychological cost: $V_\delta(h_{j}(\theta_{j}),h_{k}(\theta_{k}))$ as a function of $\theta_{j}$ and 
$\theta_{k}$, the angles between $\mathbf{J}_j$, $\mathbf{J}_k$  and  the {\it Zeitgeist} vector $\mathbf{Z}$ where $h_{j}=\cos \theta_{j}$.
 The potential can be written as $V_\delta(h_{j},h_{k})= -h_{j}h_{k}$ if  $h_{j}h_{k}<0$ (disagreement) and 
  $V_\delta(h_{j},h_{k})= \delta h_{j}h_{k}$ if  $h_{j}h_{k}>0$   (agreement). The figure depicts cases with $\delta=0$(left), 
$\delta=0.4$(center)  and $\delta=1.0$(right). 
The noisy learning dynamics tends to change the $\mathbf{J}$ making  $V_\delta$ decrease along its gradient. 
Four peaks represent the cost of maximum disagreement when moral state vectors are opposite and angles are $(0,\pm \pi)$ and $(\pm \pi,0)$. 
Note that when agents agree about the sign of their opinions, the benefit of agreement increases with $\delta$.}
\label{fig1}
\end{figure}

The functional form of the psychological cost must reflect experimental data. 
The only stylized fact we include is that there  are different cognitive styles
regarding the different way that novelty and corroborating data is handled.
By integrating the modulation function proposed in equation \ref{modulation} we reach a reasonable choice, by no means unique, that is
 depicted in figure 
\ref{fig1}:
\begin{equation}
V_\delta(h_{i},h_{j})= \frac{1}{2}(1-\delta)\rvert{h_{i}h_{j}}\lvert -\frac{1}{2} (1+\delta) h_{i}h_{j}.
 \label{psychological_cost}
\end{equation}

The corroboration/novelty parameter $\delta$ ($0\le \delta \le 1$) quantifies a cognitive strategy with respect to the difference 
in treatment of agreement and disagreement as it is suggested by ERP experiments. Our agents are conformists, namely, in the face of disagreeing opinions, the dynamics is such that the social cost is decreased. For $\delta=0$ (leftmost panel of 
figure \ref{fig1}) agents are novelty seekers and do not use corroborating opinions since $V_{\delta=0}(h_{j},h_{k})$ is flat for 
opinions of the same sign. For $\delta=1$ (rightmost  panel of figure \ref{fig1}) agents seek corroboration and conformity, learning 
equally in the case of agreement or disagreement. 

The techniques of statistical mechanics permit obtaining 
collective or aggregate emergent phenomena arising from reinforcement learning with the value of the social cost $\mathcal{H}$  or at least,
 its average, constituting relevant information to characterize the state of the society with respect to the current {\it Zeitgeist}.

\subsection{Incomplete information, statistical mechanics and peer pressure}
The incompleteness of the available information about agent
moral vectors imposes the use of probabilities in describing
the moral state of the society. What is the information available to 
construct the theory? From the evidence about cognitive styles of
 persons we have constructed two functions. First, the psychological cost,
equation \ref{psychological_cost},
which describes the cost of disagreeing
 as a function of the opinions of the agents. Second, the social cost, 
equation \ref{social_cost},
is the sum  over pairs of interacting agents
of the psychological cost. The social cost is about the whole 
society and results obtained from it will include collective 
properties arising from the interaction among the agents. It carries
two types of information, about the internal space of pair interactions
or cognitive level and about the external space, specifically about
 the geometry of the neighborhood of
interactions, or the social level. 

A complex system such  as a society can  be described in many ways.
Suppose that we choose to study experimental 
questions which will have the same
answer when the social cost has a given value or a given expected value,
averaged over the probability distribution. There might be questions
that do not fall into this category. In a physics language, an 
experimentalist wishes to prepare a system, by deciding on the control
of certain parameters, in such a way that repetitions of the experiment
will result in compatible answers for a class of questions. There will
be other questions that, not resulting in predictable answers, can't be
addressed within that experimental setup. 
This might be because, at last they are not interesting,
or that another experimental design is needed in order to examine them. 
We concentrate on those
questions for which knowing the expected value of the social cost 
is sufficient. If this cost is not known, then the most tempting
thing to do is, by claiming insufficient reason to pick one direction
over other, is to assign a uniform distribution for the 
set of moral vectors. Now, upon learning that the social 
cost of the society is an important quantity that defines the state,
we suppose it known. This simple assumption leads to the introduction
of a conjugate variable, the peer pressure scale. Knowledge of one
permits calculating the other, although this might be very difficult to do. 
At any rate, if we ignore both, for a given experimental system, the
theory requires that one of them be measured. 
This is how it goes. We start from a uniform distribution 
$P_0(\{\mathbf{J}_i\})$. Suppose that new information is obtained,
now the expected value of ${\cal H}$ is known.
This is the average with respect to  an unknown  distribution
 $P_B(\{\mathbf{J}_i\})$, which we have to find. 
Whatever was codified into the prior distribution, it was for a reason.
The new distribution will have to include the new information and in 
some sense, from all those that do, will have  to ``lie closer''  to the prior.
Closer means, effectively, that the fewest unwarranted new hypotheses must be introduced. The method to do this exists, 
and has its roots in 
Boltzmann, Gibbs, Shannon and Jaynes. See \cite{Caticha} for a modern
exposition and justification of the Maximum Entropy method.

The resulting method consists of maximizing the cross entropy between 
the prior and the posterior distributions, subject to the constraints
imposed by the new information and normalization. The constraints are
included via the usual method of Lagrange multipliers:

\begin{widetext}
\begin{eqnarray}
S\left[P_B ||P_0 \right]&=&-\int  \prod_i d\mu(\mathbf{J}_i)
P_B \ln \frac{P_B}{P_0}
+\alpha\left(E-\int \prod_i d\mu(\mathbf{J}_i){\cal H}P_B \right)
+\lambda\left(1-\int\prod_i d\mu(\mathbf{J}_i)P_B \right)
,
\end{eqnarray}
\end{widetext}
with $d\mu(\mathbf{J}_i)$ being the uniform measure on the surface of a sphere
in $M_D=5$ dimensions.

It follows that the probability of configuration $\{\mathbf{J}_i\}$ is the Boltzmann distribution
\begin{equation}
 \mathcal{P}(\{\mathbf{J}_i\}) \propto  \exp\left[-\alpha \mathcal{H}(\{\mathbf{J}_i\}) \right].
 \label{boltzmann}
\end{equation}

The Lagrange multiplier $\alpha$ is still free and has to be chosen to impose that the
average value of $\mathcal{H}$ is $E$. The informational content of $E$ and $\alpha$ is, therefore, the same. Expected values of
quantities of interest can be calculated for different values of $\alpha$ and of any parameters that enter
in $\mathcal{H}$, such as $\delta$. We name the new parameter $\alpha$ the {\it peer pressure}, since it sets the scale of the 
effect of social cost, and measures the inverse level of noise  in  the communication channel.

\section{Data on moral foundations}
Data consisted of five dimensional  score vectors with components in the interval $[0,5]$  representing the relevance attributed to each
 moral foundation. Each vector was also labeled by the subject's self-declared political affiliation from {\it p.a.}$=1$ (very liberal) 
to {\it p.a.}=$7$ (very conservative)\cite{Graham,Haidt3}.

Scores were extracted from Moral Foundations Questionnaires (MFQ30)\footnote{Available at moralfoundations.org 
(accessed on February 7, 2011).} taken by $N=14250$ US citizens. These questionnaires combine Studies 1 and 2 reported in \cite{Graham} and 
are composed by two parts each with $15$ sentences  ($3$ for each foundation) plus one verification sentence. 

In the first
 part subjects are asked  the question: 
{\it``When you decide whether something is right or wrong, to what extent are the following considerations relevant to your thinking?''}. 
Answers are given by scaling sentences of moral content from {\it``not at all relevant''} (score$=0$) to {\it``extremely relevant''} (score$=5$).  
In the second part subjects scale sentences of a moral content from  {\it``strongly disagree''}(score$=0$) to 
{\it``strongly agree''}  (score$=5$). A moral vector component is then the average of $6$ scores corresponding to a particular moral foundation.
Moral vectors $\mathbf{J}_i$ are obtained by normalizing score vectors.

\begin{figure}[th]
\includegraphics[width=9cm]{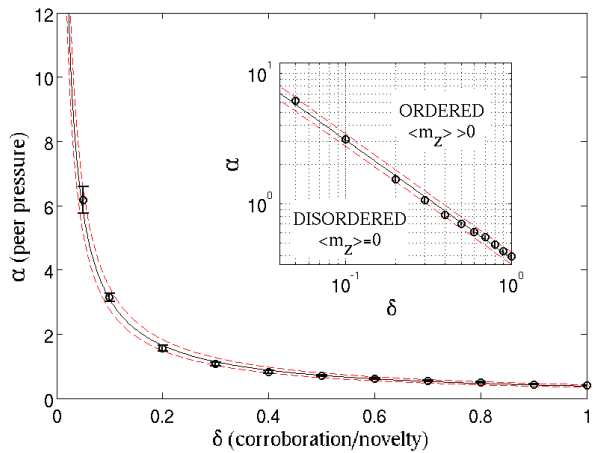}
\vspace*{8pt}
\caption{The transition line separates phases with zero (below the line) and non zero average overlap with the $\mathbf{Z}$
 vector ($\langle m_{Z_i}\rangle > 0$). The phase transition is continuous. Symbols represent average and dispersion for $20$  
simulation runs of a $N=400$ system with scale-free Barab\'asi-Albert topologies. The full line represents a fit to the 
transition border line $\alpha=k/\delta$, with $k$  constant. 
This can be seen more clearly in the inset. Dashed lines represent  $95 \%$ confidence intervals for a regression.}
\label{fig2}
\end{figure}

\section{Results}

We describe the aggregate behavior of the model and compare it to the aggregate behavior
 extracted from MFQ30 questionnaires. By introducing appropriate
 order parameters we can compare both systems, the set of subjects and the agents, in a semantic
 free manner. The correlation of political affiliation and different cognitive styles is established
 by first showing that different cognitive styles are associated to different distributions of moral 
values in the agent model and noticing that different sets of moral values are associated to different 
political affiliations in the MFQ30 data. A group of socially interacting members with a diversity of 
cognitive styles will therefore present a political spectrum. Groups of conservative agents show larger
 in-group coherence while groups of liberal agents adapt faster to changes in the issues under discussion.

\begin{figure}[th]
\includegraphics[width=9cm]{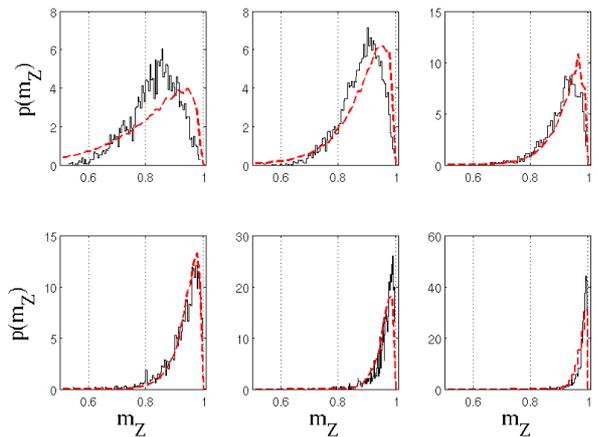}
\vspace*{8pt}
\caption{ Histograms for the effective dimension  $m_{Z_i}$ for p.a.=$1$  to $7$.
 Histograms for simulations at $\alpha=8$ and different values of $\delta$ are depicted as dashed lines for comparison. 
Simulations have been performed with a scale-free social network of size $N=400$. Results are qualitatively robust to changes 
in the lattice topology and system size. }
\label{fig3}
\end{figure}

We now discuss the statistical signatures that can be used to characterize the effective number of moral foundations of an agent.
 We compare them with equivalent signatures derived from the MFQ30 data.

Our main concern is the difference in the distribution of weights attributed to moral foundations by self-declared 
liberals and conservatives. Numerical simulation techniques, briefly discussed in the appendix, show that the model can have
 two qualitatively different regimes depending on the parameters $\delta$ and $\alpha$ (figure 2). For low $\delta$ and low $\alpha$, 
the system is in a disordered state characterized by random correlations between the moral state vectors of the agents.
 Increasing either $\delta$ or $\alpha$, a transition line can be crossed into a partially ordered society. 
Now the agents are correlated to a symmetry breaking direction $\mathbf{Z}$, the {\it Zeitgeist vector},
 which can be regarded as describing the cultural environment. The average agent is parallel to $\mathbf{Z}$. Now we reorient $\mathbf{Z}$,
 by rotating the frame of reference, so  that its components are equal (e.g. $1/\sqrt{5}$ each), explicitly assuming the equivalence of 
all moral dimensions. Note that opinions are rotation invariant and rotating makes no numerical difference. But it does foster
 interpretation, since a measure of the effective number of moral foundations of agent $i$ can be defined as proportional to the sum 
over the moral dimensions $a$, of the agents moral weights: 
\begin{equation}
 m_{Z_i}=\sum_{a=1}^5 J_{ia}Z_a,
\end{equation}
the overlap between the moral vector  and $\mathbf{Z}$, ranging from $-1$ to $1$. An agent with all moral dimensions equally important has 
$m_{Z_i}=1$. Smaller values mean it relies on a reduced subset of moral dimensions. From the survey data, we extract, for each person a 
similar measure $m_{Z_i}$ of their number of moral dimensions. 

\begin{figure}[th]
\includegraphics[width=9cm]{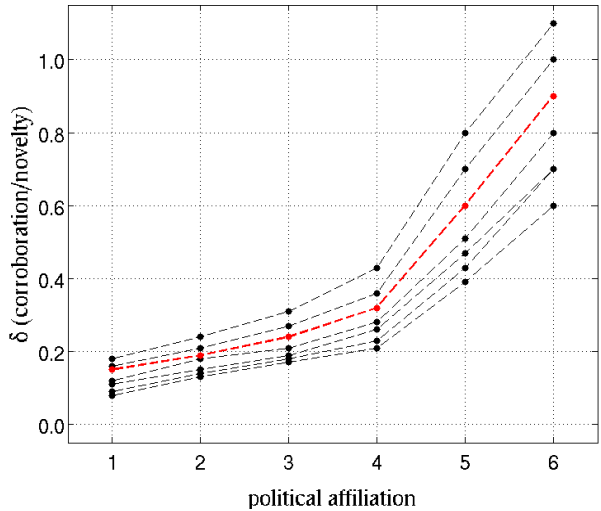}
\vspace*{8pt}
\caption{ By matching the mean average dimension, the relationship between the cognitive strategy parameter $\delta$ and 
 the political affiliation is identified for $\alpha=6$ to $12$. The dashed line represents the case depicted in figure \ref{fig3}. 
Simulations were performed with a scale-free social network of size $N=400$. Results are qualitatively robust to changes 
in the lattice topology and system size. }
\label{fig4}
\end{figure}

Our aim is to compare the statistics of $m_{Z_i}$  from the data and from the model. Figure \ref{fig3} compares histograms of $m_{Z_i}$ as 
obtained from the data and as generated by the model, for $\alpha=8$ on a scale-free social
 network. We have done several studies including different versions of the model.
The  conclusions we present, as far as the temptation of detailed quantitative confrontation 
with the data is tamed,  are independent of the different variants of 
the model. We have only considered symmetric and homogeneous interactions which allows for the use of a single 
$\delta$ throughout the social network. In all simulations presented in this paper the social neighborhood was represented by a scale-free 
random graph generated by a Barab\'asi-Albert model \cite{Barabasi} with branching rate $m=8$. 
 We have also simulated a society 
subscribed to a two dimensional 
square lattice with nearest neighbor interactions. While this neighborhood seems too artificial, the results were qualitatively 
similar to those  reported in this paper. Results within scale-free topologies, however, show the best agreement with data as far as 
overlap $m_{Z_i}$ histograms depicted in figure \ref{fig3}a are concerned.

Agents have no political affiliation and persons do not declare their cognitive strategy $\delta$. However,  histograms permit 
identifying a political affiliation with a cognitive strategy. Figure \ref{fig4} was prepared by calculating $\langle m_{Z_i} \rangle$
 for each p.a. class of the data and then finding, for each given fixed $\alpha$, the parameter 
$\delta$ that matches $\langle m_{Z_i} \rangle$  ($20$ Metropolis runs, $\alpha=6-12$). Figure \ref{fig4} also shows that the connection between
the corroboration/novelty parameter $\delta$ and political affiliation is qualitatively robust  for  a reasonably wide range of $\alpha$.

\section{Discussion}

The observations of the last section permit establishing the following link: political affiliations are partially derived from subsets of moral foundations, which arise
 collectively from distinct cognitive strategies. We conclude that the link described in the literature connecting political
 affiliation to cognitive style \cite{Graham,Haidt1} arises as a consequence of social interactions. 

As the order-disorder border line (figure \ref{fig2}) is approached from the ordered phase, the overlap with $\mathbf{Z}$ decreases, 
vanishing at the phase boundary. The best resemblance of the data and simulations occurs by identifying conservatives
 with agents far into the ordered phase and liberals with agents near the transition line but still in the ordered phase. 
Order and disorder refer to long range correlations and should not be attached to judgments of value.

We can go beyond the average number of moral foundations and make a prediction, based on the behavior of the agent model, 
about the width of the histograms. They decrease with 
increasing $\delta$ and the data shows that they decrease also with conservative tendency.
 The same identification: novelty seeking behavior to liberals, corroboration to conservatives, is again seen to arise as a 
consequence of collective behavior.

Order-disorder transitions can be driven by changing the peer pressure. Even without crossing the phase boundary, 
the model can be used to understand collective swings from left to right, as external conditions impose increased levels of peer 
pressure arising from the perception of threats. The reverse swing can also be understood when conditions demand higher adaptability
 to new challenges. We claim that with respect to moral issues, despite the differences in opinion derived from differential reliance on 
moral foundations, both conservatives and liberals are on the same side of the border. Other scenarios are discernible from the phase 
diagram. In an application outside the realm of morality, by looking at opinions on issues for which peer pressure might be lower, a
 group of large $\delta$ agents, relying on corroboration, could be found in a disordered phase and seem on this set of issues, to be liberal. 

This theory is semantically neutral. Evolutionary considerations should be used to dress the theory with semantics and to understand why 
certain foundations of morality have emerged before others and why they are different, thus breaking the remaining symmetry between the 
five dimensions. Our model cannot claim to shed light on the different nature of the different moral foundations. It just states that
 based on differential treatment of novel and corroborating information, on conformity seeking behavior and on social interactions, 
populations will present collective statistically different moral valuations in a way that can be quantitatively described. 

We believe that this work may create a number of opportunities for future research. Firstly it is highly desirable to test the model 
against new data sets. To give three illustrative examples: 1. in \cite{Fowler2} a connection is made between the structure of social networks during adolescence and political 
preferences in adulthood; 2. in \cite{Trusov} it is found that the opinion of a typical member of a virtual social network is influenced by about $20\%$ of their neighbors; 
3. \cite{Sunstein} makes an empirical analysis of the voting patterns of US federal judge panels finding correlations between the political affiliation of the majority 
and the decisions reached. Our view is that the model we have proposed might have something 
to say about what sort of patterns are expected to be seen in each one of these examples.

We suggest that a social cost can be defined and that its mean value is directly associated to a parameter $\alpha$ we have identified as the ``peer pressure''. We regard the measurement 
of peer pressure as a relevant open problem suggested by our modeling effort. Finally, the model also has some consequences that could be empirically verified, for example: 1.  peer pressure 
acts as a control parameter that if increased  can transform a group regarded 
as liberal into a conservative group. The reverse also being true; 2. a liberal group would adapt to changes in the environment faster than a conservative group but would not reach 
consensus easily.

We consider, however, as the most important contribution of this work to emphasize a particular methodological approach to the social sciences.
 From the description of how individuals react to incoming information obtained from social psychology empirical methods and neurocognitive
 data, we built an interacting model. Statistical  mechanics leads to aggregated predictions which are tested against extensive data sets with partial 
information about populations. The exchange of information and the learning it elicits, induce collective emergent properties in the 
society not to be found in the individual. Presumably it may be useful to understand how cultural divides, such as those between 
conservatives and liberals, arise partly as consequences of diversity of neurocognitive mechanisms.

\section*{Acknowledgments}
Experimental data furnished by Jonathan Haidt, whom we thank for kindly 
permitting us to have access to the data. This work received support from FAPESP
 and CNPq under grant 550981/07-1.

\bibliographystyle{ws-acs}
\bibliography{moral}

\appendix

\section{Methods summary}

\subsection{Metropolis sampling}
We assume that the model society is represented at  micro-scales by  continuous opinions $h_j$ and  that the 
system statistics can be described at intermediary time scales by a Boltzmann distribution:
\begin{equation}
P\left( \{ {\bf J}_j \} \mid  {\bf Z} \right)=\frac{1}{{\cal Z}(\alpha,\delta)} \exp \left[-\alpha\sum_{(i,j)} V_\delta(h_{i}, h_{j})\right]
\label{boltz}
\end{equation}
  with  $(i,j)$ being edges of a social graph and $V_\delta(h_{i }, h_{j})$ is the psychological cost 
 given by (\ref{psychological_cost}).
The statistics for the overlaps
  $m_{Zi}= \sum_{a=1}^{M_d}J_{ia} Z_{a}$, depicted in figure \ref{fig3}a, can be obtained by sampling from (\ref{boltz}). 
This has been done by  employing a classical Metropolis sampling technique \cite{Newman99}. 
The C code employed is provided as an ancillary file to this document.

We choose a random {\it Zeitgeist} vector ${\bf Z}$. The distribution (\ref{boltz}) is symmetric in relation to 
sign changes  $Z_{a} \rightarrow -Z_{a}$ in the components of this vector. To deal with this degeneracy all simulations 
are started with moral vectors ${\bf J}_j$ aligned to the direction ${\bf Z}$. More realistic information exchange dynamics 
to be published elsewhere shows that the system auto-organizes into the same macrostates obtained by this simplified procedure.

\subsection{Wang-Landau algorithm}
While Metropolis-like algorithms sample from the distribution and collect 
data at a single point in the phase diagram, there is another class of
algorithms which permit collecting information that will allow to obtain
results for a set of parameter values. The Wang-Landau algorithm 
\cite{Wang01}, belongs to this second class. The main theme is to 
collect information about the density of states, which in 
this case is peer pressure ($\alpha$) independent  and then to propagate
to different values of $\alpha$, 
by re-weighting via the Boltzmann factor. 
This is done for a particular value of  the novelty/corroboration 
parameter $\delta$. The density depends on $\delta$ and so this
procedure has to be repeated for a set of $\delta$ values. 
The C code employed is  provided as an ancillary file to this document.

The transition line in figure \ref{fig2} was obtained by Wang-Landau sampling
 of a system with Hamiltonian $\mathcal{H}$ at temperature $1/\alpha$ by finding numerically  
 the maximum of the specific heat for fixed $\delta$.

\subsection{Empirical histograms}
Data consisted of $N=14250$ moral vectors with components 
related to five Moral Foundations in the interval $[0,5]$ 
extracted from a specially designed the MFQ30 questionnaire\cite{Graham}. Each vector was labeled 
by the subject's self-declared political affiliation (from {\it p.a.}
$=1$ to $7$). We first calculated normalized  moral vectors   $\mathbf{J}_i$ and, by defining the  vector $\mathbf{Z}$ as 
the average vector within the conservative ({\it p.a.}$=6$) and very conservative ({\it p.a.}$=7$) classes, we have calculated histograms 
for the effective number of moral dimensions  $m_{Zi}= \sum_{a=1}^{M_d}J_{ia} Z_{a}$ (depicted in figure \ref{fig3}a). 

\begin{widetext}
\begin{table}[ht]
{\begin{tabular}{@{}cccccr@{}}\toprule
\textit{p.a. score} & n & $\langle m_z \rangle$  &  $\mu_{1/2}(m_z)$  & $\sigma_z$   & p.a. label   \\
	\colrule
1 & $2919$ & $0.825(5)$ & $0.837(4)$ & $0.084(2)$  &  very liberal                \\
2 & $5604$ & $0.877(2)$ & $0.889(2)$ & $0.069(2)$  &  liberal                     \\
3 & $2009$ & $0.907(3)$ & $0.920(3)$ & $0.063(4)$  &  slightly liberal            \\
4 & $1448$ & $0.932(3)$ & $0.947(3)$ & $0.056(4)$  &  moderate                    \\
5 & $879$  & $0.964(2)$ & $0.975(2)$ & $0.035(3)$  &  slightly conservative        \\
6 & $1087$ & $0.979(2)$ & $0.986(1)$ & $0.026(4)$  &  conservative                 \\
7 & $300$  & $0.976(4)$ & $0.987(2)$ & $0.040(10) $  &  very conservative            \\
6+7 &$1387$ &$0.979(2)$&  $0.987(1)$ & $0.028(4)$  &  conservative \\
	\botrule
\end{tabular}}
\footnotetext[1] {Error bars represent $95\%$ symmetrized bootstrap confidence intervals.}
\footnotetext[2] {$\mu_{1/2}(m_z)$ denotes the median of the overlaps $m_z$.}
\footnotetext[3] {We  consider the classes ``conservative" and ``very conservative" together 
as their statistical moments, shown in the table above, are indistinguishable.}
\label{sumario}
\end{table}
\end{widetext}

\end{document}